\documentclass[aps,prl,twocolumn,showpacs,superscriptaddress,floatfix]{revtex4}
\usepackage{epsfig}
\begin{document}

\title{Entangled light from white noise.}

\author{M.B. Plenio}
\affiliation{Quantum Optics and Laser Science Group, Blackett
Laboratory, Imperial College of Science, Technology and Medicine,
London, SW7 2BW, UK}
\author{S.F. Huelga}
\affiliation{Division of Physics and Astronomy, Department of
Physical Sciences, University of Hertfordshire, Hatfield AL10 9AB,
UK}
\date{\today}

\begin{abstract}
An atom that couples to two distinct leaky optical cavities is
driven by an external optical white noise field. We describe how
entanglement between the light fields sustained by two optical
cavities arises in such a situation. The entanglement is maximized
for intermediate values of the cavity damping rates and the
intensity of the white noise field, vanishing both for small and
for large values of these parameters and thus exhibiting a
stochastic-resonance-like behaviour. This example illustrates the
possibility of generating entanglement by exclusively incoherent
means and sheds new light on the constructive role noise may play
in certain tasks of interest for quantum information processing.
\end{abstract}
 \maketitle

Quantum entanglement holds the key for qualitatively new forms of
information processing \cite{Plenio V 98}. This discovery has
fueled an increasing interest in fully understanding how to
create, manipulate and exploit entanglement, a resource which
which has no classical analogue. From a practical point of view,
creating and exploiting entanglement has become tantamount to
minimizing the impact of noise. In any real experimental scenario,
the unavoidable interaction of the quantum processor with its
surroundings results in a decoherence process. As a consequence,
whatever initial entanglement may be present in the system of
interest, it will subsequently degrade. The system may then end up
in a mixed state whose amount of entanglement does not suffice to
overcome the performance of classically correlated states. It is
therefore of paramount importance for the practical realization of
quantum information processing protocols to engineer mechanisms to
prevent or minimize the impact of environmental noise.\\ \indent
The general approach of research so far has been to try to isolate
potential quantum information processors as much as possible from
the environment. The existing approaches can be classified into
three categories. Quantum error correction \cite{Shor} uses
redundant coding to protect quantum states against noisy
environments. This procedure is successful provided the error rate
is sufficiently small. It requires a significant overhead in
resources, both in the number of additional qubits and additional
quantum gates, which makes this technique challenging within the
limitations of presently available technology. A more economic
approach consists of exploiting the existence of so-called
decoherence-free subspaces that are completely insensitive to
specific types of noise, for example dephasing \cite{Palma}. This
approach tends to require fewer overheads, but is only applicable
in specific situations. As an example, Kielpinski et al.
\cite{nist} have recently showed how the lifetime of an entangled
pair of ions held in an ion trap can be increased by an order of
magnitude. A third theoretical approach consists of implementing
loop control strategies, where the use of ancillary systems is
avoided at the prize of interacting with the quantum processor,
either using deterministic \cite{loops} or stochastic
\cite{vitali} dynamic control. The {\em modus operandi} of all
these strategies relies on either trying to shield the system from
the environmental noise or actively restore the corrupted dynamics
to the ideal one.
\begin{figure}[htb]
 \epsfysize=6.025cm
    \begin{center}
        \epsffile{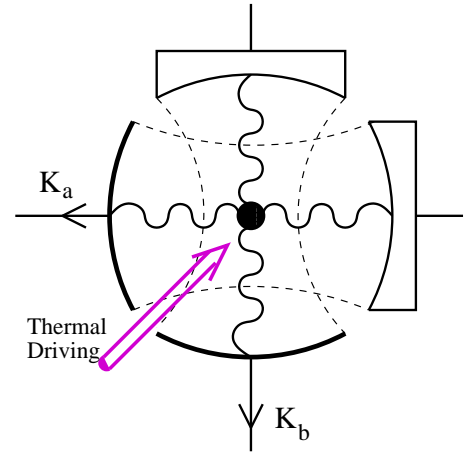}
    \end{center}
    \vspace*{-0.5cm}
\caption{\label{Fig1}Proposed experimental set up. An atomic
system is surrounded by two optical cavities initially prepared in
an arbitrary state and driven by a thermal field. Both cavities
leak photons at rates $\kappa_a$ and $\kappa_b$ . Despite the
presence exclusively of incoherent processes the steady state
solution of the system reveals that the cavity fields become
entangled.}
\end{figure}
In this article we will explore the possibility of adopting a
different strategy. Instead of attempting to combat noise, we {\em
use} noise to play a constructive role in quantum information
processing. We will focus here on the problem of generating
entanglement when {\em only} incoherent sources are available and
demonstrate that, indeed, controllable entanglement can arise in
this situation.

The idea that dissipation can assist the generation of
entanglement has been put forward recently \cite{Plenio}. In a
system comprising two atoms held in an optical cavity, it was
shown that the decay of the cavity field can be employed to assist
the preparation of a pure maximally entangled state of the ions
inside the cavity. Without cavity decay, the reduced state of the
two-ion system is an inseparable mixture at all times. Including
photon leakage leads to the undesired parts of the global
wave-function to decay and, therefore, such terms are eliminated
for sufficiently large times. This idea was then further
elaborated on to show that cavity decay can be used to establish
entanglement between different cavities \cite{Bose}.

Our approach here is more radical. We will employ white noise, not
as an exclusively passive, dissipative element, but as the actual
driving force of the system, and the only one. Consider the
situation depicted in Figure 1. An atomic system is surrounded by
two distinct optical cavities initially prepared in the vacuum
state. The system is driven by an external thermal field whose
intensity will be characterized in terms of an effective photon
number $n_T$. Both cavities may leak photons at rates $\kappa_a$
and $\kappa_b$.
The Bloch equations governing the time evolution of the global
system are given by (in the following $\hbar=1$)
\begin{equation}
    \dot\rho = -i \left[ H,\rho \right] + {\cal L}(\rho),
    \label{Master equation}
\end{equation}
where the Hamiltonian $H$ describes the internal energies of the
atom and the two cavities as well as the atom-cavity coupling. The
Liouvillean ${\cal L}(\rho)$ describes the cavity decay and the
interaction of the atom with the external thermal light field. As
no external coherent driving is present, the Hamiltonian reads
\begin{eqnarray}
    H &=& \nu_a a^{\dagger}a  + \nu_b b^{\dagger}b
    +  \omega |2\rangle\langle 2 |\nonumber \\
    && \!\!\!\!\!\!\!\!\!\!\!\! +  g_a
    (|2\rangle\langle 1|a + |1\rangle\langle 2|a^{\dagger}) + g_b
    (|2\rangle\langle 1|b + |1\rangle\langle 2|b^{\dagger})
\end{eqnarray}
where $g_{a(b)}$ is the coupling constant of the atom to cavity
modes $a(b)$ and $|i\rangle$ denotes the internal state $i$ of the
atom. The Liouvillean is given by
\begin{eqnarray}
    {\cal L}(\rho) &=&
    - \kappa_a \left[a^{\dagger}a\rho + \rho a^{\dagger}a - 2 a \rho a^{\dagger}\right]
    \nonumber \\
    && - \kappa_b \left[b^{\dagger}b\rho + \rho b^{\dagger}b - 2 b \rho
    b^{\dagger}\right]\nonumber \\
    &&-(n_T+1)\Gamma \left[|2\rangle\langle 2| \rho + \rho |2\rangle\langle 2| -
    2 |1\rangle\langle 2|\rho|2\rangle\langle 1|\right]\nonumber\\
    && - n_T\Gamma \left[|1\rangle\langle 1| \rho + \rho |1\rangle\langle 1| -
    2 |2\rangle\langle 1|\rho|1\rangle\langle 2|\right] \; .
\end{eqnarray}
Here $\Gamma$ describes the coupling strength of the atom to the
external fields and $n_T \Gamma$ is the transition rate due to the
thermal field. The spectral width of the thermal field is large
compared to the linewidth of the atomic transition so that its
effect is that of a white noise source. Note that $n_T$ can be
interpreted as an effective photon number and that spontaneous
decay of the atom out of the cavities is included in this scenario
via the $n_T+1$ term. To simplify the following considerations, we
assume that $\nu_A=\nu_B=\omega$ and chose an interaction picture
with respect to $H_0 = \omega (|2\rangle\langle 2| + a^{\dagger}a
+ b^{\dagger} b)$. Using this transformation, the Liouvillean part
remains unchanged, while the Hamiltonian part is now given by
\begin{eqnarray}
    H_I =  g_a (|2\rangle\langle 1|a + |1\rangle\langle 2|a^{\dagger})
    + g_b (|2\rangle\langle 1|b + |1\rangle\langle
    2|b^{\dagger})\; .
\end{eqnarray}
The general solution of the equation (\ref{Master equation}) is
extremely tedious. Therefore we study a special case that exhibits
the typical behaviour but which allows to reduce the complexity,
both analytical and numerical, of the problem considerably. This
is the case where both cavities have the same decay rate, ie we
assume $\kappa_a=\kappa_b$.

In this case we are able to introduce two new, effective modes one
of which will be decoupled from the atom. This simple basis change
is given by the definition of two new mode operators
\begin{equation}
    c=\frac{g_a a + g_b b}{\sqrt{g_a^2 + g_b^2}} \,\,\,\,\, \mbox{and}
    \,\,\,\,\,
    d=\frac{g_b a - g_a b}{\sqrt{g_a^2 + g_b^2}} \, .
\end{equation}
and corresponds to a beam-splitter transformation. In this new
basis, the Hamiltonian and Liouvillean part of the master equation
are given by
\begin{equation}
    H_I =  g \left( c |2\rangle\langle 1| + c^{\dagger} |1\rangle\langle
    2|\right), \label{eq5}
\end{equation}
where $g=\sqrt{g_a^2+g_b^2}$, and with $\kappa=\kappa_a=\kappa_b$,
\begin{eqnarray}
    {\cal L}(\rho) &=&
    - \kappa \left[c^{\dagger}c\rho + \rho c^{\dagger}c - 2 c \rho
    c^{\dagger} + d^{\dagger}d\rho + \rho d^{\dagger}d - 2 d \rho d^{\dagger}\right]\nonumber \\
    && -(n_T+1)\Gamma \left[|2\rangle\langle 2| \rho + \rho |2\rangle\langle 2| -
    2 |1\rangle\langle 2|\rho|2\rangle\langle 1|\right]\nonumber \\
    &&- n_T\Gamma \left[|1\rangle\langle 1| \rho + \rho |1\rangle\langle 1| -
    2 |2\rangle\langle 1|\rho|1\rangle\langle 2|\right]. \label{eq6}
\end{eqnarray}
It is straightforward to transform between the different mode
pictures. For example given a Fock state of the effective cavity
modes $c$ and $d$, we can express it in terms of a linear
combination of Fock states of the physical modes $a$ and $b$ via
%\begin{eqnarray}
    $\sqrt{m! n!}|m\rangle_c|n\rangle_d = \left(c^{\dagger}\right)^n
    \left(d^{\dagger}\right)^m |vacuum\rangle$.
%    &&\hspace*{-1.5cm} \frac{1}{\sqrt{m! n!}}
%    \left(\frac{g_a a^{\dagger} + g_b b^{\dagger}}{\sqrt{g_a^2 + g_b^2}}\right)^{m}
%    \left(\frac{g_b a^{\dagger} - g_a b^{\dagger}}{\sqrt{g_a^2 +
%    g_b^2}}\right)^{n}|vacuum\rangle.
%\end{eqnarray}
We observe that, due to the transformation to the new set of
effective modes, we have one mode (mode $d$) which is completely
decoupled from the Hamiltonian dynamics and is purely damped under
the Liouvillean dynamics. As a consequence, irrespective of the
initial state of mode $d$, it will not be populated in steady
state. Therefore, we begin all our numerical investigations with
both effective modes $c$ and $d$ in the vacuum state (which is
equivalent to both physical modes $a$ and $b$ being in the ground
state). As the mode $d$ will then never be populated, we disregard
mode $d$ in the following. With this approach both the analytical
and the numerical integration becomes much more efficient and we
can now study the entanglement physics of our system.
Nevertheless, the analytical expression for general parameter
values are extremely tedious \cite{mathematica} and we begin with
the presentation of some numerical results and then provide a
physical explanation for them.

The typical behaviour of the entanglement in the system is
illustrated in Figure $2$. There we have plotted the amount of
entanglement of the joint state of the two cavity modes (after
tracing out the atomic system) as a two-variable function of the
intensity of the thermal field (characterized by the effective
photon number $n_T$) and time. The chosen parameters are
$\nu_a=\nu_b=\omega$, $g_a=g_b=1$, $\kappa_a=\kappa_b=1$ and
$\Gamma=0.2$. This entanglement implies non-classical correlations
between the modes which are a resource in quantum communication
tasks \cite{Plenio V 98} or the violation of Bell inequalities
\cite{Hardy 98}. It is quantified by the so-called logarithmic
negativity $N(\rho)$, which is given by
\begin{equation}
   N(\rho) = \log ||\rho^{\Gamma_B}||
    \label{negativity}
\end{equation}
where $\rho^{\Gamma_B}$ is the partial transpose of $\rho$ and
$||\rho^{\Gamma_B}||$ denotes the trace norm of $\rho^{\Gamma_B}$,
which is the sum of the singular values of $\rho^{\Gamma_B}$
\cite{Werner 01}. We have chosen the logarithmic negativity as it
has the advantage that it can be computed fairly easily for
systems of arbitrary dimensions, in contrast to other established
entanglement measures such as entanglement of formation or the
relative entropy of entanglement \cite{otherentmeasures}.
$N(\rho)$ is non-negative and vanishes for states with positive
partial transpose and, in particular, for separable states.
\begin{figure}[htb]
 \vspace*{-1.75cm}
 \epsfysize=11.cm
 \epsfxsize=9.cm
    \begin{center}
        \epsffile{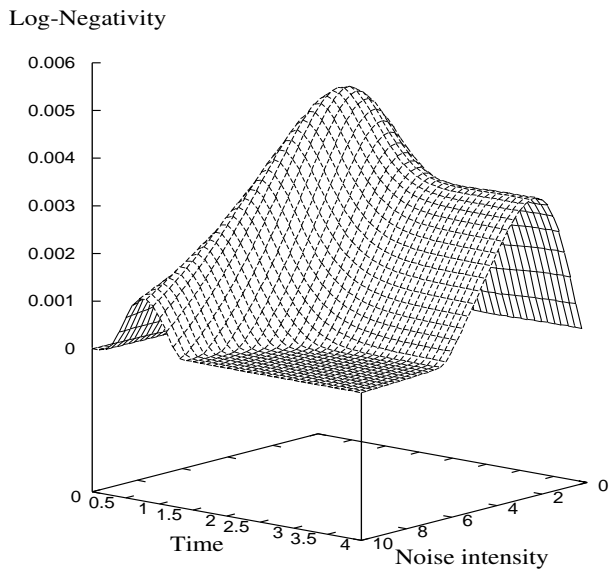}
    \end{center}
    \vspace*{-2.cm}
\caption{\label{fig2} Amount of entanglement $N(\rho)$, defined by
eq. (\protect\ref{negativity}), of the two-cavity field state as a
function of the intensity of the external incoherent driving
(characterized by the effective photon number $n_T$) and the time
at which we determine the state of the light fields. We have
chosen $\nu_a=\nu_b=\omega$, $g_a=g_b=1$, $\kappa_a=\kappa_b=2$
and $\Gamma=0.2$. Note that the entanglement achieves its maximum
for an intermediate intensity of the noisy driving field, a
phenomenon reminiscent of stochastic resonance
\protect\cite{Hangi}.}
\end{figure}
In our simulation we cut off the intra-cavity photon number at a
value of $3$ (simulations with a cutoff of $4$ photons lead to
indistinguishable results). Note that for any value of $t$, the
behaviour of the amount of entanglement between the cavity modes
is non-monotonic, it increases to a maximum value for an optimal
intensity of the noisy driving field and then decreases towards
zero for a sufficiently large intensity. This behaviour is
reminiscient of the well-known phenomenon of stochastic resonance
\cite{Hangi} (see also \cite{srquant}) where the coherent response
of a system to a periodic signal is maximized in the presence of
an intermediate amount of noise.
\begin{figure}[htb]
 \vspace*{-1.75cm}
 \epsfysize=11cm
 \epsfxsize=9.cm
    \begin{center}
        \hspace*{0.4cm}\epsffile{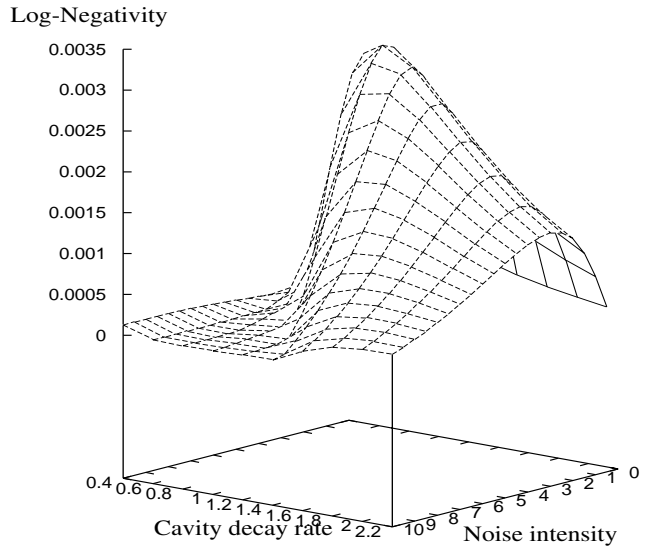}
    \end{center}
    \vspace*{-2.cm}
\caption{\label{fig3}The steady state entanglement of the two
cavity modes as a function of both the cavity decay rate and the
intensity of the noisy driving field. The parameters are
$\nu_a=\nu_b=\omega$, $g_a=g_b=1$ and $\kappa=\kappa_a=\kappa_b$.
The system exhibits a resonance for both parameters. Entanglement
only persists when both incoherent processes are present with
moderate values. }
\end{figure}
This first approach where the atom is traced out may appear rather
crude and a more sophisticated approach would involve a
measurement on the atom. Indeed, the expected amount of
entanglement between the light fields can be increased if we
subject the atomic system to a projective measurement in the
$\{|1\rangle,|2\rangle\}$ basis. We have not plotted the resulting
curves as they exhibit qualitatively the same behaviour as in
Figure $2$ but with an entanglement that is approximately twice as
large.

It is also worthwhile to study the steady state entanglement of
the cavity light fields as a function of both the intensity of the
noisy driving field and of the cavity decay rates. In Figure $3$
we present these result for $\nu_a=\nu_b=\omega$, $g_a=g_b=1$,
$\kappa=\kappa_a=\kappa_b$ and $\Gamma=0.2$. The system exhibits a
double resonance in both parameters again reminding of stochastic
resonance. This demonstrates that both the noisy driving field and
the cavity decay are necessary ingredients to generate steady
state entanglement in this system. Note that a related effect
assisted by cavity decay has been found recently by Nha et al.
\cite{korea} when studying the generation of squeezing in a
coherently driven atom-cavity system. There, the squeezing is
maximized as the cavity decay is increased to an optimal value and
degrades for larger decay rates.

To understand the origin of the generation of entanglement from
white noise and cavity decay let us begin by considering the
special case of perfect cavities, ie $\kappa=0$. One easily checks
that the stationary state of the joint (atom-cavity field) system
in the effective mode picture is then given by
\begin{eqnarray}
    \rho^{ss} &=&
    \sum_{r=0}^{\infty} \sum_{i=1}^{2} \left(\frac{n_T}{1+n_T}\right)^{r+i-1}
    \frac{|i\rangle_A\langle i|\otimes |r\rangle_C\langle r |}{2n_T +1} \label{effective}
\end{eqnarray}
where $|r\rangle_C=(c^{\dagger})^r|vacuum\rangle/\sqrt{r!}$,
$|i\rangle_A$ denotes the internal atomic states and $n_T$ is
defined as in eq. (3). This can also be written as
\begin{eqnarray}
    \!\rho^{ss} \!\!\!&=&\!\!
    \sum_{r=0}^{\infty} \sum_{i=1}^{2}
    \left(\frac{n_T}{1+n_T}\right)^{r+i-1}\!
    \frac{|i\rangle_A\langle i|\otimes |\psi^{ab}_r\rangle\langle \psi^{ab}_r |}{2n_T+1} \label{original}
\end{eqnarray}
where
$|\psi^{ab}_r\rangle=(a^{\dagger}+b^{\dagger})^r|vacuum\rangle/||.||$.
Tracing out the atomic state results in a thermal distribution for
the two cavity modes. Note that a thermal state of mode $c$ is
also a mixture of coherent states and that the transformation back
into the original two-mode picture (modes $a$ and $b$) corresponds
to a beam-splitter transformation. A thermal state impinging on a
beam splitter leads to a separable two-mode state after the
beam-splitter and we conclude without further calculations that
the state in eq. (\ref{original}) is not entangled.

If we increase the value of the cavity decay constant for fixed
thermal noise intensity $n_T$, the stationary state is no longer a
thermal mixture. For very large $\kappa$ only the lowest energy
levels of the modes are occupied (the population of the effective
mode $c$ with $r$ photons decreases as $(1/n_T)^{2+3(r-1)}$ for
$r\ge 1$). A mixture of the vacuum and the one photon state in the
effective mode picture transforms into a mixture of vacuum and the
triplet state $(|01\rangle+|10\rangle)/\sqrt{2}$. Such a mixture
is always entangled as one easily confirm by computing its partial
transpose \cite{otherentmeasures}. Therefore we expect that in the
presence of cavity damping some entanglement will be present. This
entanglement will again vanish in the $\kappa\rightarrow\infty$
limit as then the stationary state tends to the vacuum state.

To shed some more light on how exactly does the cavity decay
generate the entanglement consider the decay of a photon in the
effective mode picture. Such a decay leads to the application of
the operator $c$ to the state. In the original mode picture this
corresponds to the application of the operator $a+b$. If this
operator is applied to the state and the state is renormalized,
then one observes that the relative weight of the entangled states
in the mixture has increased. The conditional time evolution if
there is no photon leakage has just the opposite effect, the mean
number of photons decreases, ie the weight of entangled states in
the mixture reduces \cite{Plenio K98}.

To summarize, we have presented a preparation technique to
entangle two light fields using only incoherent processes.
Entanglement is present in the system at all times and its exact
amount is a non monotonic function of the intensity of the
external, incoherent driving. This behaviour resembles the
phenomenon of stochastic resonance, where the response of a
non-linear system to a weak periodic driving can be enhanced when
supplemented with a noisy field of certain optimal intensity
\cite{Hangi} (See \cite{srquant} for proposed demonstrations in
the quantum domain). In our case, the amount of entanglement of
the two light fields is maximized for an optimal value of the
external noisy field. If the intensity of the incoherent driving
is increased beyond this optimal value, the negativity of the
state decreases. This situation exemplifies the fact that tunable
noisy sources can play a constructive role in certain situations
and opens a new venue for exploring efficient ways to exploit the
presence of noise when the aim is to generate entanglement in a
controllable way.

{\em Acknowledgements:} We acknowledge comments from participants
of the ESF-QIT conference
%'Quantum Information: Theory, Experiment and Perspectives'
in Gdansk, July 2001, where the main lines of this work were
presented. This work was supported by EPSRC, the Nuffield
Foundation, the EQUIP project of the European Union, the European
Science Foundation Programme on 'Quantum Information Theory and
Quantum Computing' and the US Army Research Office grant
43371-PH-QC.


\begin{thebibliography}{99}
%
\bibitem{Plenio V 98} M.A.\ Nielsen and I.L.\ Chuang,
{\it Quantum Computation and Quantum Information}\/ (Cambridge
University Press, Cambridge, 2000); M.B. Plenio and V. Vedral,
Contemp. Phys. {\bf 39}, 431 (1998).
%
\bibitem{Shor} P.W. Shor, Phys. Rev. A {\bf 52}, 2493 (1995); A.R. Calderbank
and P.W. Shor, Phys. Rev. A {\bf 54}, 1098 (1996); A.M. Steane,
Proc. Roy. Soc. A{\bf 452}, 2551 (1996).
%
\bibitem{Palma} G.M. Palma {\em et al},
%, K-A Suominen and A.K. Ekert,
Proc. Roy. Soc. A{\bf 452}, 567 (1996); M.B. Plenio {\em et al},
%,V. Vedral and P.L. Knight,
Phys. Rev. A {\bf 55}, 67 (1997); D.A. Lidar {\em et al},
%, I.L.Chuang and K.B. Whaley,
Phys. Rev. Lett. {\bf 81}, 2594 (1998).
%
\bibitem{nist} D. Kielpinski {\em et al},
%, V. Meyer, M. A. Rowe, C. A. Sackett, W. M. Itano, C. Monroe, and D. J. Wineland,
Science {\bf 291}, 1013 (2001).
%
\bibitem{loops}H.M. Wiseman and G.J. Milburn, Phys. Rev. Lett. {\bf 70}, 548
(1993);  D. Vitali {\em et al},
%, P. Tombesi and G.J. Milburn,
Phys. Rev. Lett. {\bf 79}, 2442 (1997); L. Viola {\em et al},
%, E. Knill and S. Lloyd,
Phys. Rev. Lett. {\bf 82}, 2417 (1999).
%
\bibitem{vitali} S. Mancini {\em et al},
%, D. Vitali, R. Bonifacio and P. Tombesi,
quant-ph/0108011.
%
\bibitem{Plenio} M.B. Plenio {\em et al},
%, S.F. Huelga, A. Beige, and P.L. Knight.
Phys. Rev. A {\bf 59}, 2468 (1999);
%A. Beige {\em et al},
%, D. Braun, and P. L. Knight,
%New J. Phys. {\bf 2}, 22 (2000);
A. Beige {\em et al},
%, S. Bose, D. Braun, S.F. Huelga, P.L. Knight, M. B. Plenio, and V. Vedral,
J. Mod. Opt. {\bf 47}, 2583 (2000); P. Horodecki, Phys. Rev. A
{\bf 63}, 022108 (2001).
%
\bibitem{Bose} S. Bose {\em et al},
%, P.L. Knight, M.B. Plenio and V. Vedral.
Phys. Rev. Lett. {\bf 83}, 5158 (1999).
%
\bibitem{mathematica} A Mathemtica program is available from
the authors
%(m.plenio@ic.ac.uk or s.f.huelga@herts.ac.uk)
that provides analytical expressions.
%
\bibitem{Hardy 98} L. Hardy, Contemp. Phys. {\bf 39}, 419 (1998).
%
\bibitem{Werner 01} G. Vidal and R.F. Werner, Lanl e-print
quant-ph/0102117; J. Eisert, PhD thesis Potsdam 2001.
%
\bibitem{otherentmeasures} C.H. Bennett {\em et al},
%, D.P. DiVincenzo, J.A. Smolin, and W.K. Wootters,
Phys. Rev. A {\bf 53}, 3824 (1996); V. Vedral and M.B.
Plenio, Phys. Rev. A {\bf 57}, 1619 (1998).
%
\bibitem{korea} H. Nha {\em et al},
%, Y-T. Chough, S.W. Kim and K. An,
quant-ph/0106054.
%
\bibitem{Hangi} L. Gammaitoni {\em et al},
%, P. H{\"a}nggi, P. Jung and F. Marchesoni,
Rev. Mod. Phys. {\bf 70}, 223 (1998).
%
%
\bibitem{Plenio K98} M.B. Plenio and P.L. Knight, Rev. Mod. Phys.
{\bf 70}, 101 (1998).
%
\bibitem{srquant} S.F. Huelga and M.B. Plenio, Phys. Rev. A {\bf 62} 052111
(2000); A. Buchleitner and R.N. Mantegna, Phys. Rev. Lett. {\bf
80}, 3932 (1998)
%; T. Wellens and A. Buchleitner, Phys. Rev. Lett. {\bf 84}, 5118 (2000).


\end{thebibliography}
\end{document}